\documentclass[aps,prc,twocolumn,superscriptaddress,floatfix,longbibliography,10pt]{revtex4-1}

\usepackage{graphicx}
\usepackage{hyperref}
\usepackage{float} 
\usepackage{xcolor}
\usepackage{amsmath}
\usepackage{amssymb} 
\usepackage[T1]{fontenc}
\usepackage[english]{babel}
\usepackage{subcaption}
\usepackage{textcomp,gensymb}
\usepackage{pifont} 
\usepackage{orcidlink}

\begin{document}


\title{Resonant Cancellation Effect in Ramsey-type Nuclear Magnetic Resonance Experiments}



\author{Ivo~Schulthess\orcidlink{0000-0002-5621-2462}}
\email[Corresponding author: ]{ivo.schulthess@unibe.ch}
\altaffiliation[Present address: ]{Deutsches Elektronen-Synchrotron DESY, 22607 Hamburg, Germany}
\affiliation{Laboratory for High Energy Physics and Albert Einstein Center for Fundamental Physics, University of Bern, 3012 Bern, Switzerland}

\author{Ivan~Calic\orcidlink{0009-0001-4552-9886}}
\altaffiliation[Present address: ]{WSL Institute for Snow and Avalanche Research SLF, 7260 Davos Dorf, Switzerland}
\affiliation{Laboratory for High Energy Physics and Albert Einstein Center for Fundamental Physics, University of Bern, 3012 Bern, Switzerland}

\author{Estelle~Chanel}
\altaffiliation[Present address: ]{Present address: Institut Laue-Langevin, CS 20156, 38042 Grenoble Cedex 9, France.}
\affiliation{Laboratory for High Energy Physics and Albert Einstein Center for Fundamental Physics, University of Bern, 3012 Bern, Switzerland}

\author{Anastasio~Fratangelo\orcidlink{0000-0001-9964-601X}}
\affiliation{Laboratory for High Energy Physics and Albert Einstein Center for Fundamental Physics, University of Bern, 3012 Bern, Switzerland}

\author{Philipp~Heil\orcidlink{0000-0001-5309-5988}}
\affiliation{Laboratory for High Energy Physics and Albert Einstein Center for Fundamental Physics, University of Bern, 3012 Bern, Switzerland}

\author{Christine~Klauser\orcidlink{0000-0002-5266-4488}}
\affiliation{Paul Scherrer Institut, 5232 Villigen PSI, Switzerland}

\author{Gjon~Markaj}
\affiliation{Laboratory for High Energy Physics and Albert Einstein Center for Fundamental Physics, University of Bern, 3012 Bern, Switzerland}

\author{Marc~Persoz\orcidlink{0009-0002-6748-6837}}
\affiliation{Laboratory for High Energy Physics and Albert Einstein Center for Fundamental Physics, University of Bern, 3012 Bern, Switzerland}

\author{Ciro~Pistillo\orcidlink{0000-0001-8131-9440}}
\affiliation{Laboratory for High Energy Physics and Albert Einstein Center for Fundamental Physics, University of Bern, 3012 Bern, Switzerland}

\author{Jacob~Thorne\orcidlink{0000-0002-3905-5549}}
\affiliation{Laboratory for High Energy Physics and Albert Einstein Center for Fundamental Physics, University of Bern, 3012 Bern, Switzerland}

\author{Florian~M.~Piegsa\orcidlink{0000-0002-4393-1054}}
\email[Corresponding author: ]{florian.piegsa@unibe.ch}
\affiliation{Laboratory for High Energy Physics and Albert Einstein Center for Fundamental Physics, University of Bern, 3012 Bern, Switzerland}


\date{\today}

\begin{abstract}
We investigate the response of a Ramsey-type experiment on an additional oscillating magnetic field. This superimposed field is oriented in the same direction as the static main magnetic field and causes a modulation of the original Larmor spin precession frequency. The observable magnitude of this modulation reduces at higher frequencies of the oscillating field. It disappears completely if the interaction time of the particles matches the oscillation period, which we call resonant cancellation. We present an analytical approach that describes the effect and compare it to a measurement using a monochromatic cold neutron beam. 
\end{abstract}


\maketitle


\section{\label{sec:intro}Introduction}

Rabi's nuclear magnetic resonance method~\cite{rabi_molecular_1939, kellogg_magnetic_1939} and Ramsey's techniques of separated oscillatory fields~\cite{ramsey_new_1949,ramsey_molecular_1950,ramsey_neutron_1986} have been used very effectively in various scientific experiments. These methods utilize both constant and time-varying magnetic fields to manipulate the spin of probe particles. Specifically, Ramsey's technique allows for precisely determining of the Larmor precession frequency of a spin in a magnetic field $B_0$. This is achieved by initially flipping the spin-polarized particles into the plane orthogonal to $B_0$ using an oscillating field $B_1$, allowing them to precess for a defined period of time, and then flipping them again using a second oscillating $B_1$ field. By scanning the frequency of the oscillating fields close to the resonance and keeping the phase between the two signals locked, an interference pattern of the spin polarization in the frequency domain is obtained. Another option is to scan the relative phase between the two oscillating spin-flip signals while keeping their frequencies on resonance. Ramsey's technique is highly versatile and can be applied for precise measurements of changes in magnetic and pseudo-magnetic fields. For instance, it is and has been utilized in a variety of applications such as atomic clocks~\cite{essen_atomic_1955,ramsey_history_1983,wynands_atomic_2005}, the measurement of the Newtonian gravitational constant~\cite{rosi_precision_2014}, in gravitational resonance spectroscopy measurements~\cite{abele_ramseys_2010, jenke_gravity_2014, bosina_qbounce_2022}, the measurement of the neutron magnetic moment~\cite{greene_measurement_1979}, the measurement of neutron incoherent scattering length~\cite{roubeau_systematic_1974,abragam_spin-dependent_1975, glattli_experimental_1979, malinovski_spin-dependent_1981, piegsa_ramsey_2008}, the search for a neutron electric dipole moment~\cite{abel_measurement_2020, piegsa_new_2013,chupp_electric_2019, kirch_search_2020, ayres_design_2021}, probing for dark matter~\cite{abel_search_2017,schulthess_new_2022}, and the search for new particles and interactions~\cite{piegsa_limits_2012}.

The main magnetic field $B_0$ that causes the Larmor precession in a Ramsey-type experiment is usually constant over time. In this article, we describe the effect of a superimposed oscillating (pseudo-)magnetic field on the Ramsey signal, which we call \textit{resonant cancellation effect}. 
Such a signal is potentially present in an axionlike dark-matter field that couples to the electric dipole moment of particles. Two experiments searching for an oscillating electric dipole moment of neutrons~\cite{schulthess_new_2022,schulthess_search_2022} and electrons~\cite{roussy_experimental_2021} showed how this systematic effect attenuates the signal and reduces their sensitivity. 
Moreover, ultralight scalar or pseudo-scalar dark matter could lead to oscillating nuclear charge radii that could be detected with Ramsey's technique in optical atomic clocks~\cite{banerjee_oscillating_2023}. Such measurements would also potentially suffer from the effect. 

In the following, we provide a theoretical derivation of the resonant cancellation effect for a monochromatic particle beam with a velocity $v$. In a Ramsey-type experiment, the spins precess in the main magnetic field $B_0$ due to the Larmor precession
\begin{equation}\label{eq:larmorPhase}
    \varphi = \omega T = \gamma B_0 T \ ,
\end{equation}
where $\gamma$ is the gyromagnetic ratio and $T$ is the interaction time. The phase shift that the spins acquire when interacting with an oscillating magnetic field ${B_a(t) = B_a \cos{(2 \pi f_a t)}}$ parallel to $B_0$ is  
\begin{equation}
    \Delta \varphi =  \gamma \int B_a \cos{(2 \pi f_a t)} \, \mathrm{d}t \ ,
\end{equation}
where $B_a$ is the amplitude and $f_a$ the frequency of the oscillating field. This becomes maximal when the integral is evaluated symmetrically around zero which leads to a maximum value of the phase shift
\begin{align}
    \Delta \varphi_\text{max} &= \left| \gamma B_a \int_{-T/2}^{T/2} \cos{(2 \pi f_a t)} \, \mathrm{d}t  \right| \\ 
    &= \left| \frac{\gamma B_a}{\pi f_a} \sin{\left(\pi f_a \frac{L}{v}\right)} \right| \ , \label{eq:resonantCancellationPhase}
\end{align}
where the interaction length $L = v T$. Equation~(\ref{eq:resonantCancellationPhase}) has the shape of a sinc-function with roots at ${f_a = k \cdot v / L}$, $k \in \mathbb{N}$. Hence, when the interaction time with the oscillating field matches one or multiple periods of the oscillating magnetic field, the integral becomes zero, yielding no phase shift. Therefore, a Ramsey-type experiment is insensitive to magnetic field effects of specific frequencies. In the case of a white beam, e.g., with a Maxwell-Boltzmann-like velocity distribution, this effect is suppressed, and there is no complete cancellation. It still results in a reduced average phase shift and, therefore, a decreased sensitivity to oscillating magnetic field effects at higher frequencies.

\section{\label{sec:setup}Experimental Setup}

\begin{figure}[!t]
    \includegraphics[width=0.48\textwidth]{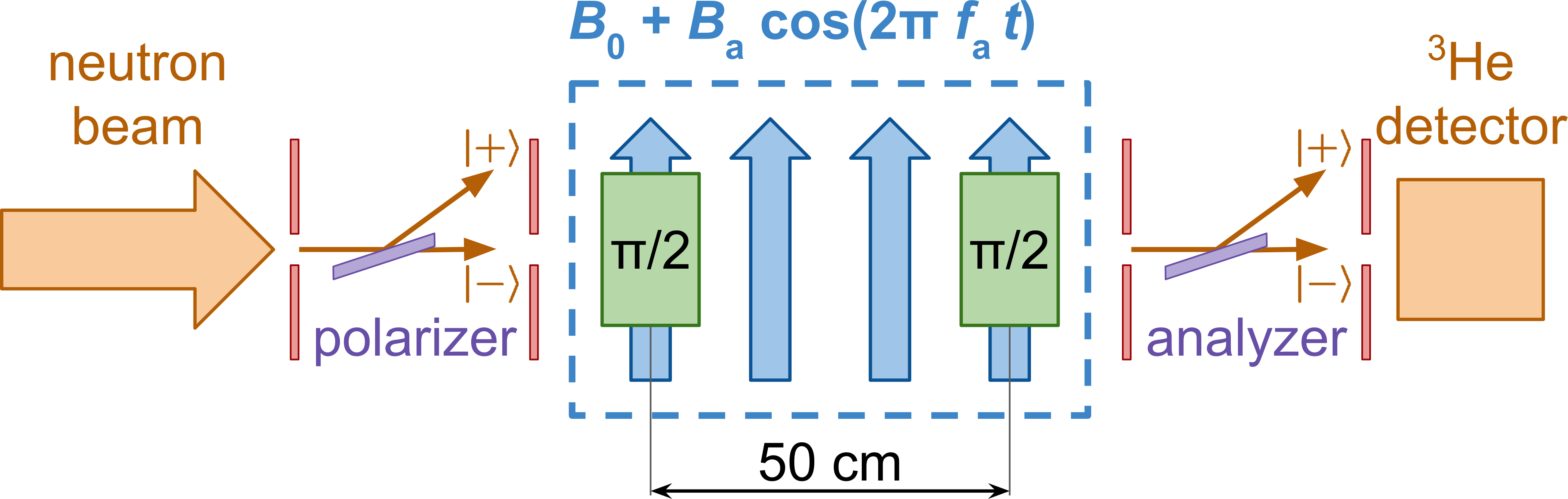}
    \caption{Schematic of the experimental apparatus. Neutrons enter the setup from the left. They first pass a polarizer that transmits only one spin state. The neutrons then enter the magnetic field region where a constant magnetic field $B_0$ and an oscillating magnetic field $B_a(t)$ are applied in the vertical direction. Two $6~\mathrm{cm}$-long spin-flip coils with a center-to-center separation of $50~\mathrm{cm}$ allow to flip the spins. After the second spin-flip coil the neutron spin states are analyzed and counted with a $^\text{3}$He detector. There are four apertures to define the beam cross-section and divergence. They also block the reflected beams from the polarizer and analyzer. }
    \label{fig:schematic}
\end{figure}

To investigate the resonant cancellation effect we conducted measurements during a dedicated beamtime in May 2021 at the Narziss instrument at the Paul Scherrer Institute in Switzerland. The Narziss beamline is usually used for neutron reflectivity measurements. It provides neutrons with a de~Broglie wavelength of $4.96~\text{\r{A}}$ and can be regarded as monochromatic for the purpose of this measurement. A schematic of the experiment is presented in Fig.~\ref{fig:schematic}. 
The neutrons pass a beam monitor after exiting from the neutron guide. This monitor signal can be used to normalize all measurements to the same number of incoming neutrons and accounts for potential fluctuations in the neutron flux of the source. A measurement is stopped when a defined number of monitor counts is reached. 
There are four apertures with an opening of $30 \times 2~\mathrm{mm^2}$ that define the beam cross-section and divergence along the beam path. Their position is indicated in Fig.~\ref{fig:schematic}. 
The neutrons are polarized with a polarizing supermirror that transmits only one spin state. The other spin state is reflected and stopped in an aperture. To maintain the polarization, there is a small magnetic guiding field between the polarizer and the first spin-flip coil. 
The main magnetic field $B_0 \approx 3~\mathrm{mT}$ is created by an electromagnet surrounding the back of a C-shaped iron yoke. The yoke guides the field lines and creates a field between its flanks in the vertical direction. 
The oscillating magnetic field $B_a(t)$ is applied over the same region as the $B_0$ field. It is created by a Keysight waveform generator~\cite{keysight_technologies_keysight_2021} that is connected via a Kepco bipolar power amplifier~\cite{kepco_inc_kepcos_2011} to an auxiliary coil. This coil can also be used to apply DC offset fields to test the functionality of the setup. 
Two rectangular solenoid-type spin-flip coils induce resonant $\pi/2$-flips. They have a length of $6~\mathrm{cm}$ and a center-to-center separation of $50~\mathrm{cm}$. Their axis is aligned with the neutron beam and their cross-section is much bigger than the size of the beam. Their signals are produced by a waveform generator and amplified by two Mini-Circuits LYZ-22+ power amplifiers~\cite{mini-circuits_coaxial_2013}. They are connected to the spin-flip coils via resonance circuits that match $50~\Omega$ and are tuned to the resonance frequency of about 90~kHz. Their two sides are covered with $2~\mathrm{mm}$ thick aluminum plates to minimize their fringe fields. 
The spins are analyzed with a second polarizing supermirror. 
The transmitted neutrons are counted using a $^\text{3}$He detector. Each neutron creates an electronic pulse. The timing of the pulse is processed and recorded by a custom-made data-acquisition system using an Arduino. 

\section{\label{sec:measurements}Measurements}

The spin-flip signals were optimized for Ramsey-type measurements. We individually measured a Rabi
resonance frequency of ${(91.94 \pm 0.01)~\mathrm{kHz}}$ and ${(91.46 \pm 0.01)~\mathrm{kHz}}$ for the first and the second spin-flip coil, respectively. For further measurements we applied a signal at a frequency of $91.7~\mathrm{kHz}$. The amplitudes of both spin-flip signals were optimized for the highest signal visibility in a Ramsey measurement, corresponding to a $\pi/2$-flip. 

\begin{figure}[!t]
  \begin{subfigure}[b]{0.48\textwidth}
    \includegraphics[width=\textwidth]{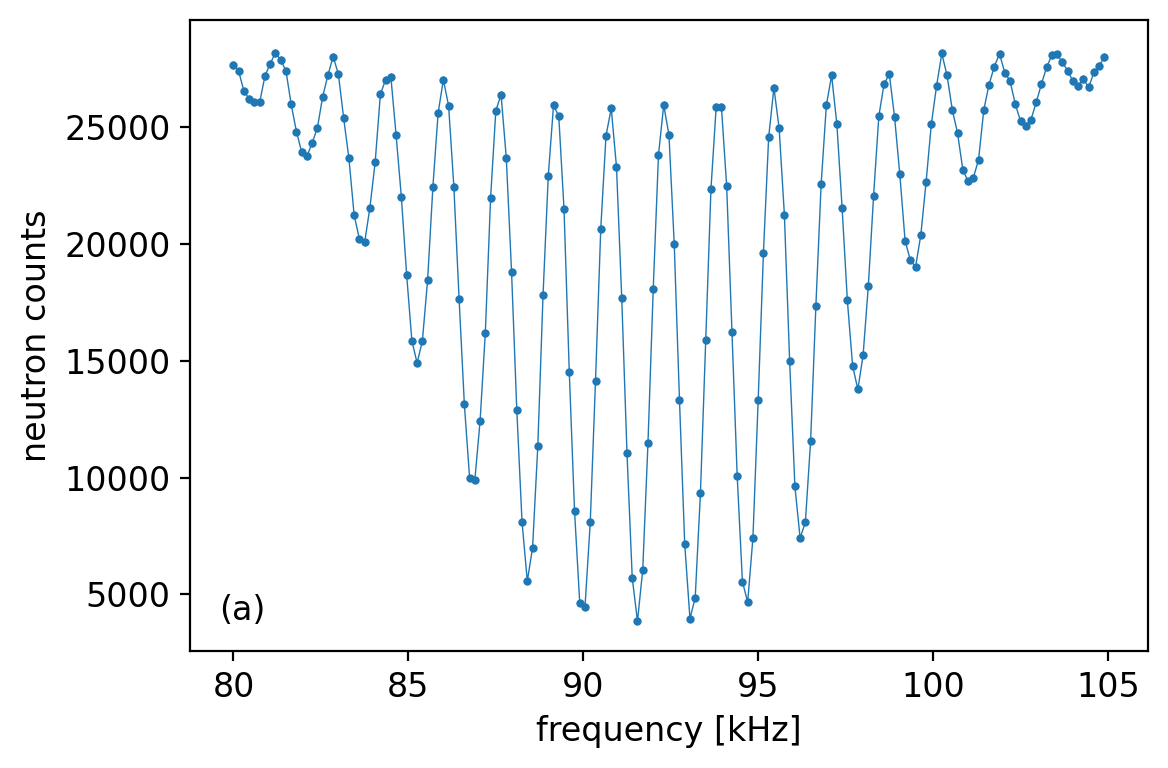}
  \end{subfigure}
  \hfill
  \begin{subfigure}[b]{0.48\textwidth}
    \includegraphics[width=\textwidth]{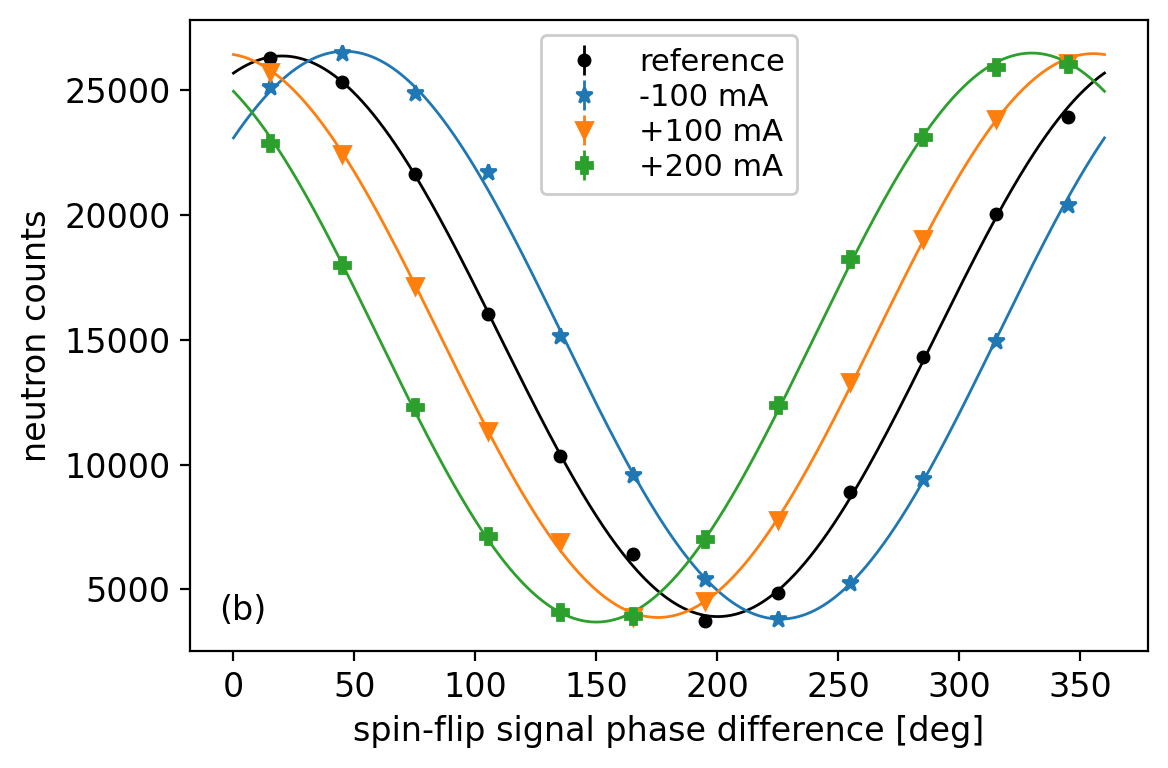}
  \end{subfigure}
  \caption{(a) Ramsey frequency scan over the full resonance. The data shows the neutron counts as a function of the spin-flip signal frequency. The solid line serves only as a guide for the eye. (b) Ramsey phase scans were the neutron counts are shown as a function of the relative phase between the two spin-flip signals. The frequency was fixed on resonance at $91.7~\mathrm{kHz}$. Besides a reference measurement (\ding{108}), we applied various DC offset currents via the auxiliary coil. The measurements with a current of $-100~\mathrm{mA}$ (\textcolor{blue}{\ding{72}}), $+100~\mathrm{mA}$ (\textcolor{orange}{\ding{116}}), and $+200~\mathrm{mA}$ (\textcolor{green}{\ding{58}}) are shown. The solid lines correspond to least-squares fits of a sinusoidal function. }
  \label{fig:ramseyMeasurements}
\end{figure}

To test the functionality and characterize the setup we performed various Ramsey frequency and phase scans. Two such measurements are presented in Fig.~\ref{fig:ramseyMeasurements}. For each data point, the neutron counts were integrated over roughly $10$~seconds, corresponding to ${2 \times 10^5}$ monitor counts. 
The Ramsey frequency scan presented in Fig.~\ref{fig:ramseyMeasurements}a shows an overall envelope which arises from the Rabi resonance. The fringes are the interference pattern produced by the two spatially separated spin-flip coils. All fringe maxima are close to the maximum neutron counts of about $2.8 \times 10^4$ neutrons, confirming the small wavelength spread. The Ramsey fringes have a visibility of roughly 76\%. 
We measured the shift of the resonance frequency by scanning only the central fringe while applying various DC offset currents through the auxiliary coil. A linear fit through all resonance frequencies results in a value for the shift of $(-1.096 \pm 0.002)~\mathrm{kHz / A}$ which corresponds to roughly $38~\mathrm{\mu T / A}$. 
The Ramsey phase scan in Fig.~\ref{fig:ramseyMeasurements}b shows a sinusoidal behavior whose phase shifts if an additional offset current is applied, as in the case of the central resonance fringe. We performed several more measurements to get the phase shift as a function of the applied current. A linear least-squares fit through all phases results in a value of $(-257.0 \pm 0.4)\mathrm{\degree / A}$. The measurements presented in Fig.~\ref{fig:ramseyMeasurements} show that the apparatus works as expected and that an offset current through the auxiliary coil indeed changes the precession frequency of the neutrons. 
The results of the Ramsey frequency and phase scans can be used to calculate the effective interaction length with the use of Eq.~(\ref{eq:larmorPhase}) and ${v \approx 798~\mathrm{m/s}}$. The resulting value of ${(51.9 \pm 0.1)~\mathrm{cm}}$ is slightly longer than the separation of the two spin-flip coils. The reason is that the spins already start to precess within the spin-flip coil when partially flipped.

To investigate the resonant cancellation effect, we measured the neutron counts continuously and saved their timing information. Because of limitations from the data-acquisition system, we phase-wrapped the data into two periods of the oscillating magnetic field $B_a(t)$. This wrapping was triggered by the waveform generator which created the oscillating signal. We chose a time bin size of $20~\mathrm{\mu s}$. This size allowed us to have more than ten bins for the highest frequency of $3500~\mathrm{Hz}$. 
The relative phase between the spin-flip signals was set to $105\degree$ which corresponds to the point of steepest slope in the reference measurement of the Ramsey phase scan shown in Fig.~\ref{fig:ramseyMeasurements}b. At this point, the measurement is most sensitive to magnetic field changes, and the relation between the neutron counts and the phase is approximately linear. 

\begin{figure}[!t]
    \includegraphics[width=0.48\textwidth]{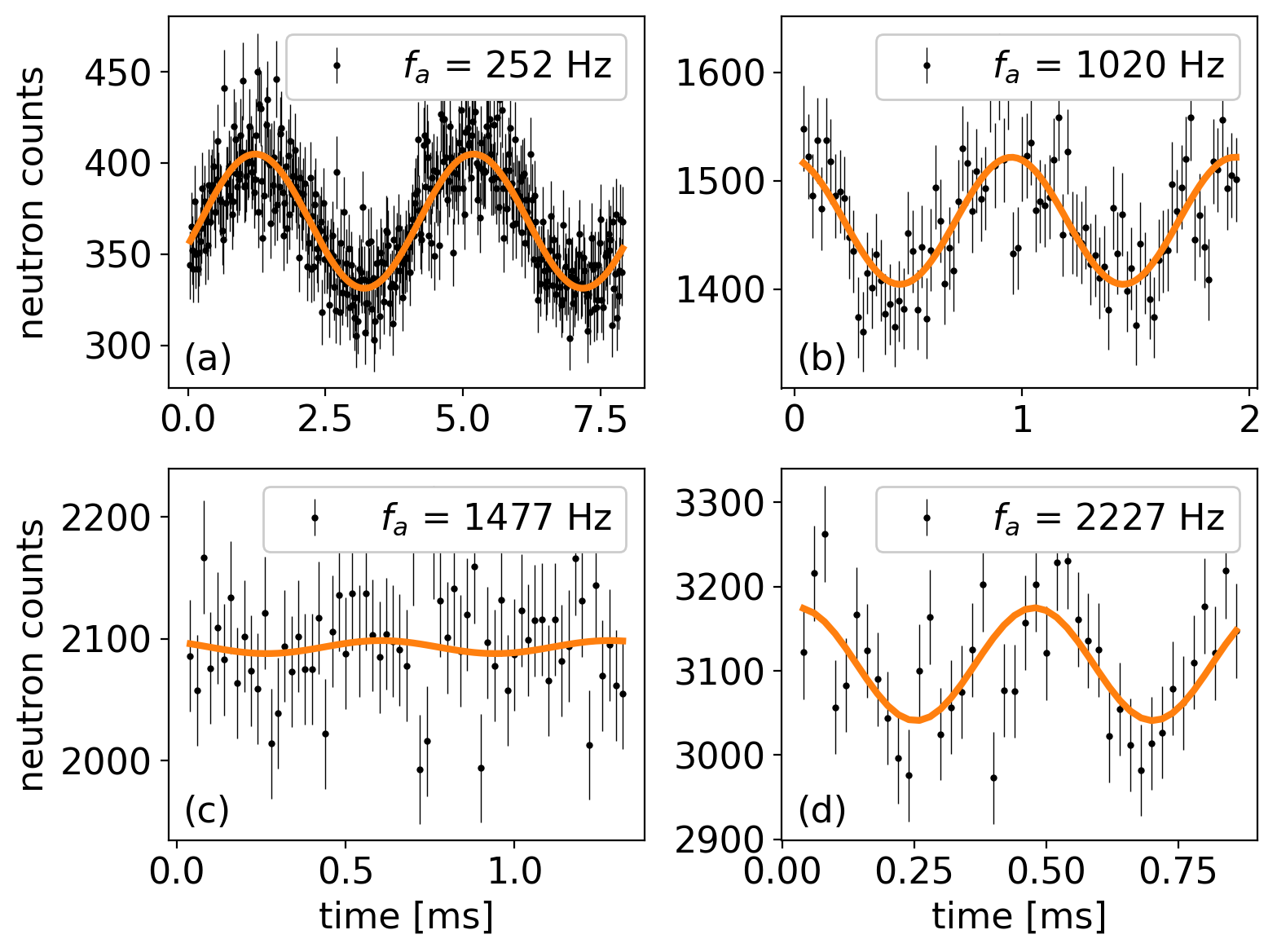}
    \caption{Neutron signals for oscillating magnetic fields at $f_a$ equal to (a) 252~Hz, (b) 1020~Hz, (c) 1477~Hz, and (d) 2227~Hz. They show the neutron counts as a function of time. The counts are phase-wrapped into two periods of the signal. The data (black) are fitted with a sinusoidal function with a fixed frequency (orange). The fit results are presented in Tab.~\ref{tab:neutronSignalFitResults}. The measurement time was roughly $100$~seconds for each setting.}
    \label{fig:neutronSignal}
\end{figure} 
\begin{table}[!t]
    \begin{tabular*}{0.48\textwidth}{@{\extracolsep{\fill}} c | c c c c} 
        $f_a$ & amplitude & offset & amplitude/offset & $\chi^2$/NDF\\
        \hline
        252~Hz & $37 \pm 1$ & $368 \pm 1$ & $0.100 \pm 0.004$ &  329/391 \\ 
        1020~Hz & $59 \pm 6$ & $1463 \pm 4$ & $0.040 \pm 0.004$ & 120/92 \\
        1477~Hz & $5 \pm 8$ & $2093 \pm 6$ & $0.003 \pm 0.004$ & 52/62 \\
        2227~Hz & $67 \pm 12$ & $3108 \pm 9$ & $0.022 \pm 0.004$ & 47/39 \\
    \end{tabular*}
    \caption{Results of the sinusoidal fit of the data presented in Fig.~\ref{fig:neutronSignal}. The amplitude is divided by the offset to make the measurements comparable. The phase of the sinusoidal fit is not relevant for the analysis of the resonant cancellation effect. The $\chi^2$ of each fit and the corresponding number of degrees of freedom are also given. }
    \label{tab:neutronSignalFitResults}
\end{table}

We measured the oscillating neutron amplitude for $100$~frequencies $f_a$ between $60~\mathrm{Hz}$ and $3500~\mathrm{Hz}$. Four examples of neutron signals are shown in Fig.~\ref{fig:neutronSignal}. The results of their sinusoidal fits are presented in Tab.~\ref{tab:neutronSignalFitResults}. 
All measurements have different periods but the same time bin sizes. Therefore, the number of time bins is different. Since the total number of neutrons is the same for all measurements, the amplitude and offset of the sinusoidal neutron signal scale with the frequency of the oscillating field. To account for this, we divided the amplitude of the neutron signal by its offset. 
To improve the signal-to-noise ratio we repeated this sequence three times and averaged the data. 
The frequency dependence of the oscillating magnetic field amplitude $B_a$ was investigated in an auxiliary measurement with a fluxgate sensor. 
It features a nearly linear decrease of roughly 13\% from 200~Hz to 2~kHz which is due to the experimental setup. The measurement was limited to this frequency range due to the bandwidth of the fluxgate. We used a fit of the fluxgate data in this range which was extrapolated to the full extend of the measurement to account for this characteristics.  
Besides the compensation of the frequency-dependent magnetic field, we normalized the ratio of amplitude over offset to one at DC. 

\begin{figure}[!t]
    \includegraphics[width=0.48\textwidth]{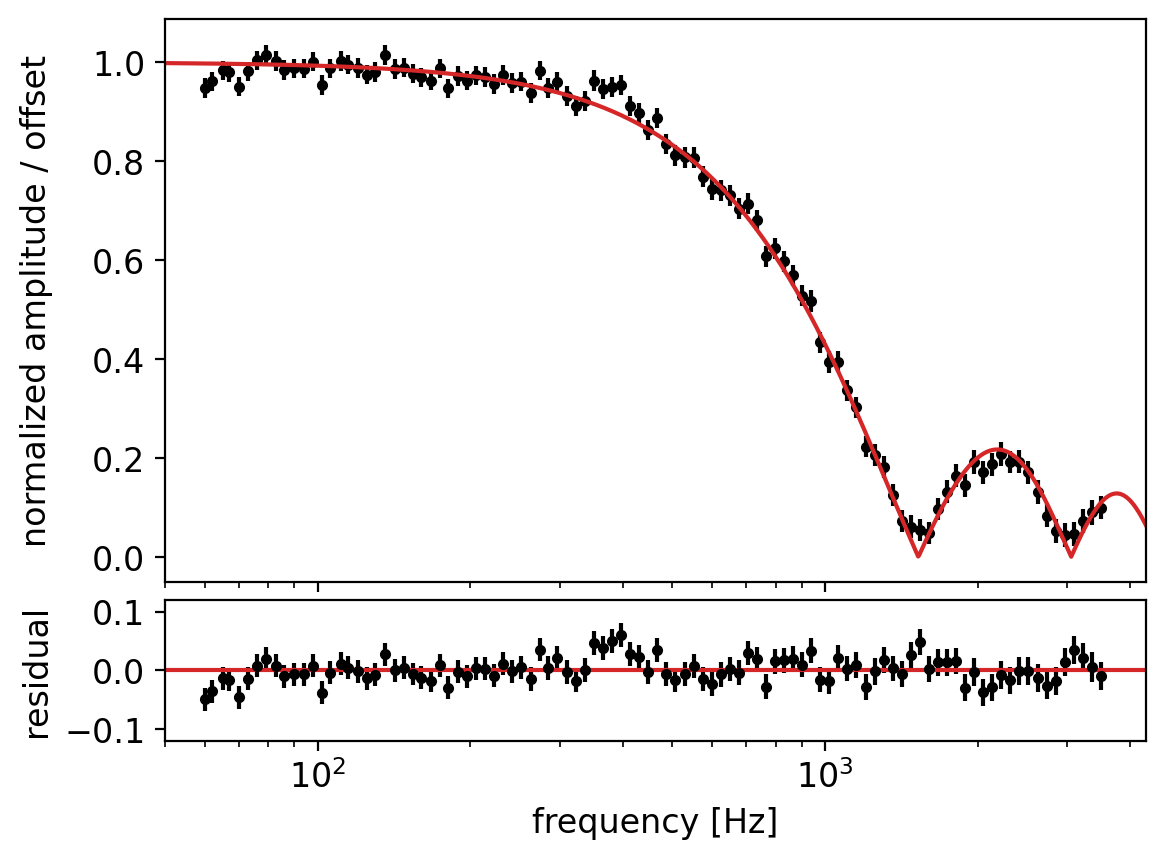}
    \caption{Measurement of the resonant cancellation effect at the Narziss beamline. The normalized ratio of amplitude over offset as a function of the frequency of the oscillating signal is shown. The data (black) are fitted with Eq.~(\ref{eq:resonantCancellationPhase}) (red). The lower subfigure shows the residuals. }
    \label{fig:resonantCancellation}
\end{figure}

The result of this measurement and analysis is shown in Fig.~\ref{fig:resonantCancellation}. 
A fit of Eq.~(\ref{eq:resonantCancellationPhase}) yields the values of the roots at $(1529 \pm 7)~\mathrm{Hz}$ and $(3057 \pm 14)~\mathrm{Hz}$. The reduced chi-square of the fit is $\chi_r^2 = 1.04$ for 98 degrees of freedom. 
Figure~\ref{fig:resonantCancellation} shows that the effect of the resonant cancellation behaves as expected for a monochromatic neutron beam. Given a neutron wavelength (velocity) of $4.96~\text{\r{A}}$ ($798~\mathrm{m/s}$), the fit values of the root leads to an interaction length of ${(52.2 \pm 0.2)~\mathrm{cm}}$. This value is in agreement with the aforementioned effective interaction length of ${(51.9 \pm 0.1)~\mathrm{cm}}$ which was determined from the Ramsey data.

\section{\label{sec:conclusion} Conclusion}
In conclusion, we investigated the resonant cancellation effect. This effect is present in Ramsey-type experiments where an oscillating magnetic field is superimposed to the main magnetic field that causes the Larmor precession. 
We showed that the neutron oscillation amplitude follows the expected theoretical behavior in the case of a monochromatic neutron beam. This results in a reduced sensitivity of the experiment for higher frequencies to the superimposed field. In particular, the sensitivity becomes zero if the interaction time of the neutrons matches the period of the oscillating field. This approach can be used to estimate the attenuation effect for experiments searching for time dependent signals. 
For instance, this systematic effect is important in axionlike dark-matter searches using electric dipole moments of particles.

\begin{acknowledgments}
We gratefully acknowledge the excellent technical support by R.~H\"anni, J.~Christen, and L.~Meier from the University of Bern. The experiment has been performed at the Swiss Spallation Neutron Source SINQ at the Paul Scherrer Institute in Villigen, Switzerland. This work was supported via the European Research Council under the ERC Grant Agreement no.\ 715031 (BEAM-EDM) and via the Swiss National Science Foundation under grants no.\ PP00P2-163663 and 200021-181996.
\end{acknowledgments}

\section*{Conflict of Interest}
The authors have no conflicts to disclose.

\section*{Author Contributions}
\textbf{I.~Schulthess:} conceptualization (equal); data curation (lead); formal analysis (equal); investigation (equal); methodology (lead); software (lead); validation (equal); visualization (lead); writing - original draft (lead); writing – review \& editing (equal). 
\textbf{I.~Calic:} formal analysis (equal); investigation (equal); software (supporting); validation (equal); writing – review \& editing (supporting). 
\textbf{E.~Chanel:} writing – review \& editing (supporting). 
\textbf{A.~Fratangelo:} investigation (equal); writing – review \& editing (supporting). 
\textbf{P.~Heil:} writing – review \& editing (supporting). 
\textbf{Ch.~Klauser:} resources (supporting); writing – review \& editing (supporting). 
\textbf{G.~Markaj:} writing – review \& editing (supporting). 
\textbf{M.~Persoz:} writing – review \& editing (supporting). 
\textbf{C.~Pistillo:} writing – review \& editing (supporting). 
\textbf{J.~Thorne:} writing – review \& editing (supporting). 
\textbf{F.~M.~Piegsa:} conceptualization (equal); funding acquisition (lead); investigation (equal); project administration (lead); resources (lead); supervision (lead); validation (equal); writing – review \& editing (equal). 

\section*{Data Availability}
The data and analysis that support the findings of this study are openly available in a Github repository~\cite{schulthess_ivoschulthessresonantcancellation_2023}.

\bibliography{references.bib}

\end{document}